\documentclass[12pt]{article}

\usepackage{epsfig}

\usepackage{amssymb}
\usepackage{amsfonts}

\usepackage{color}

\usepackage{hyperref}

\def\be{\begin{equation}}
\def\ee{\end{equation}}
\def\ba{\begin{array}{c}}
\def\ea{\end{array}}

\newcommand{\bea}{\begin{eqnarray}}
\newcommand{\eea}{\end{eqnarray}}

\def\ssk{\medskip}
\def\bsk{\bigskip}
\def \la {\label}
\def \pa {\partial}
\def \fr {\frac}
\def \f {\phi}
\def \y {\psi}
\def \ve {\varepsilon}
\newcommand{\rf}[1]{(\ref{#1})}

\begin{document}

\vspace{2.9cm}

 \begin{center}{\Large \bf

Applied Supersymmetry

  }

\vspace{2cm}

  {\bf V.P. Berezovoj}$^{a}$ and {\bf A.J. Nurmagambetov}$^{a,b,c}$\footnote{ajn@kipt.kharkov.ua; a.j.nurmagambetov@gmail.com}

\end{center}

\vspace{3mm}

 $^{a}${Akhiezer Institute for Theoretical Physics of NSC KIPT, 1 Akademichna St., Kharkiv 61108, Ukraine}

 $^{b}${Department of Physics \& Technology, Karazin Kharkiv National University, 4 Svobody Sq., Kharkiv 61022, Ukraine}

 $^{c}${Usikov Institute of Radiophysics and Electronics, 12 Ak. Proskury, Kharkiv 61085, Ukraine}

\vspace{5mm}



%
\abstract{
Supersymmetry is one of the most important and indispensable ingredients of modern theoretical physics. However, the absence, at least at the time of publishing this review, of experimental verification of supersymmetry in elementary particle/high-energy physics casts doubt on the viability of this concept at the energies achievable at the LHC. The more unexpected are either already experimentally verified or proposed for verification manifestations of supersymmetry at the level of low (condensed matter physics, quantum optics) and medium (nuclear physics) energies where standard quantum mechanics works. Using examples of various systems from completely different areas of physics, we discuss the isospectrality of quantum Hamiltonians, hidden and explicit supersymmetry, the advantages of the supersymmetric quantum mechanics approach and its role in quantum engineering.
}

\bsk\bsk

\ssk

\newpage

\section{Introduction}

On July 3, 2025, we celebrate 100 years since the birth of Dmitry Volkov, academician of the National Academy of Sciences of Ukraine, the founder of one of the scientific schools of the Kharkov Institute of Physics and Technology, one of the inspirers of the creation of the O.I. Akhiezer Institute for Theoretical Physics of NSC KIPT. During his 45-year activity at UFTI-KFTI-NSC KIPT \cite{VolkovBook} Volkov made a number of discoveries that completely overturned our understanding of fundamental physics. Among them are: the Green-Volkov-Greenberg-Messiah parastatistics \cite{Green:Para53,Volkov:Para59,Greenberg:Para64,Messiah:Para64}; Volkov's (jointly with V.N. Gribov) discovery of the ``Regge Poles Conspiracy'' \cite{Gribov:1963gx} and the subsequent development of dual models (see the review \cite{Volkov:DM1975} and references therein); phenomenological Lagrangians of fundamental interactions \cite{Volkov:1973vd} (and Refs. therein); supersymmetry \cite{Golfand:1971iw,Volkov:1972jx,Volkov:1973ix,Wess:1973kz,Wess:1974tw}; supergravity \cite{Volkov:1973jd,Deser:1976eh,Freedman:1976xh}; supersymmetric quantum mechanics \cite{Nicolai:1976xp,Witten:1981nf,Witten:1982im,Volkov:SQM1985,Volkov:1986xe}, spontaneous compactification of extra dimensions \cite{Volkov:1980kq}, twistor approach in supersymmetric theory of particles \cite{Volkov:1989ct,Sorokin:1988jor,Sorokin:1988nj}, strings \cite{Volkov:1990vk} and branes \cite{Bandos:1995zw,Bandos:1995dw}. Volkov's scientific school members also played an important role in the above-mentioned works, but, of course, he himself set the tone for the research.

Among the above-mentioned scientific inventions of D.V. Volkov, perhaps the main one is the discovery and development of Supersymmetry -- a symmetry of a new type, unifying in {\it supermultiplets} particles of the same quantum numbers, but different statistics. In the end of 60th it was already known that a non-trivial unification of space-time symmetries with gauge symmetries is forbidden within the formalism of the standard quantum field theory \cite{McGlinn:1964ji,Coleman:1967ad}. Supersymmetry (the historical retrospective of the discovery and development of supersymmetry is given in \cite{Kane:2000ew}) was one of the tools that allowed to overcome this obstacle, which finally led to the appearance of supergravity as a unified theory of space-time and internal (gauge) symmetries. Being a part (to be precise, a low-energy limit) of string theory, the development of supergravity made it possible to take a different look at the Dual Models of strong interactions, and to make two superstring revolutions on their basis (see \cite{StringTheory} for a brief review). At present string theory, transformed into M-theory, is considered the most probable candidate for building a unified theory of all fundamental interactions. Although, certainly, the program of construction of the Theory of Everything is still far from its final completion.

It is usually accepted to consider that supersymmetry, growing from particle physics, has a relation to (ultra)high energies, beyond the reachable energies of modern particle colliders. Indeed, the renormalizability property of supersymmetric theories immediately attracted the attention of theorists in the field of high energy physics. But the absence of experimental evidence of supersymmetry on the LHC energy scale clearly indicates a strong violation of supersymmetry. The idea about spontaneous supersymmetry breaking on scales of TeVs is not new. In order to investigate the properties of the supersymmetric vacuum and spontaneous supersymmetry breaking Witten \cite{Witten:1981nf,Witten:1982im} initiated the development of supersymmetric quantum mechanics \cite{Nicolai:1976xp}, which was a significant step for the application of supersymmetry in low- and medium-energy physics.

Nowadays Supersymmetry \cite{Golfand:1971iw,Volkov:1972jx,Volkov:1973ix,Wess:1973kz,Wess:1974tw} plays an important role not only in the world of elementary particle physics, but also in nuclear, solid states and condensed matter physics, quantum mechanics, quantum optics and many other areas of contemporary physics. In this brief review we are aimed at pointing out some features of supersymmetry in the description of various physical systems, which favourably distinguish this approach from the conventional ones.
To recap supersymmetric quantum mechanics (SQM), the approach we promote to describe quantum objects, in Section 2 we recall main points of the construction of isospectral quantum-mechanical Hamiltonians within the standard and extended schemes. In Section 3 we give examples of models from different areas of physics in which supersymmetry is hidden. On the contrary, in Section 4 we consider the situation with exact supersymmetry of stationary Hamiltonians and outcomes arising therefrom. The recently obtained non-trivial connection of CPT-invariant stationary and PT-invariant non-stationary supersymmetric Hamiltonians is discussed in Section 5. Application of the SQM technique for open quantum systems with mixed states and the Lindbladian dynamics of the density matrix is the subject of Section 6. Conclusions contain a brief discussion on the results and further directions of studies.


\section{A brief recap of Supersymmetric Quantum Mechanics }

Let’s refine main steps in constructing supersymmetric Hamiltonians. We can follow the factorization method proposed by Dirac and Schr\"odinger in the mid of 20th of the last century. For the sake of simplicity, we will apply the Dirac-Schr\"odinger formalism to one-dimensional stationary Hamiltonians, when generally the Hamiltonian operator is presented by
\be
H_0=-\frac{d^2}{dx^2}+V_0\left(x\right).
\label{H0}
\ee                         
According to the formalism, one presents the initial Hamiltonian \rf{H0} as
\be
H_0=A^\dag A+\epsilon,
\la{H0AA}
\ee                           
with an analog of the creation/annihilation operators
\be
A=\frac{d}{dx}+\beta\left(x\right),\qquad A^\dag=-\frac{d}{dx}+\beta\left(x\right).         
\la{AAdef}
\ee
The unknown function $\beta(x)$, entering eq. \rf{AAdef}, can be found from the Riccati equation
\be
-\frac{d}{dx}\beta\left(x\right)+\beta^2\left(x\right)=V_0\left(x\right)-\epsilon. 
\la{Ric}
\ee             
Though this construction is general, and can be applied to any quantum-mechanical Hamiltonian, the general solution to the Riccati equation is absent in the analytic form. Put it differently, one cannot resolve eq. \rf{Ric} for an arbitrary potential $V_0(x)$, as well as the Schrodinger equation with Hamiltonian \rf{H0}. 

However, there is a set of quantum-mechanical potentials, for which the exact solutions to the Riccati equation exist. These potentials are referred to as exactly-solvable potentials of Quantum Mechanics (QM). And in this case, we can apply the factorization scheme and determine the operators \rf{AAdef} and the factorization energy $\epsilon$ explicitly \cite{Infeld:1951mw}. 

But it is not the end of the story. In 1984 Mielnik \cite{Mielnik:1984jmp} proposed the way of constructing a new Hamiltonian from the original one, if a particular solution to the Riccati equation has known (see \cite{Kampen:1971}). If we denote this particular solution as $\beta_0(x)$, then the new Hamiltonian will receive the structure of
\be
H_1=AA^\dag+\epsilon=-\frac{d^2}{dx^2}+V_1(x)
\la{H1}
\ee
with the new potential
\be
V_1\left(x\right)=V_0\left(x\right)+2\frac{d}{dx}\beta_0(x) .
\la{V1}
\ee
The new potential can be absolutely different in shape. However, the spectra of Hamiltonians $H_0$ and $H_1$ turn out to be related to each other via the so-called intertwining relations:
\be
 H_1A=AH_0,\qquad H_0A^\dag=A^\dag H_1 .
\la{intertwin}
\ee

When the starting Hamiltonian $H_0$ is an exactly-solvable one, we can choose 
\be
\beta_0=-\frac{\varphi_\epsilon^\prime(x)}{\varphi_\epsilon(x)} ,
\la{partSol}
\ee
and the still arbitrary factorization energy $\epsilon$ becomes a part of the spectrum of one of the paired Hamiltonians. Specifically, as $\epsilon<E_0$ (where $E_0$ is the ground state energy of $H_0$), the factorization energy is the ground state energy of the new Hamiltonian $H_1$. As a result, the intertwined Hamiltonians $H_0$ and $H_1$ are (almost) isospectral; their spectra are different just in the ground energy state (see Figure 1).

\begin{figure}[ht]
\begin{center}
\includegraphics[width=2.5in]{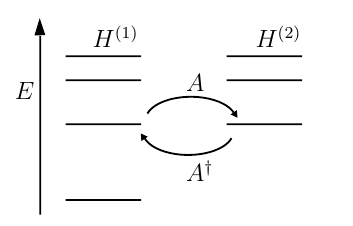}
  \caption{Schematic spectra of two intertwined Hamiltonians related by (7). Here $H^{(1)}\equiv H_1$, $H^{(2)}\equiv H_0$.}\label{Fig1}
  \end{center}
\end{figure}

So far, we did not even mention Supersymmetry. However, there is a very close connection \cite{Witten:1981nf,Krive:1985ufn} between operators $A$ and $A^\dag$ of \rf{AAdef} and the so-called supercharges $Q$ and $Q^\dag$ forming the part of the minimal Supersymmetry algebra:\footnote{We will return to the specifics of Supersymmetric Quantum Mechanics in Section 4 in more detail.}
\be
\left\{Q,Q^\dag\right\}=H .
\la{SAlg}
\ee
Then, the paired from the point of view of relations \rf{intertwin} Hamiltonians form the so-called {\it supermultiplet}. Since properties of a (super)multiplet members have to be the same, it is not surprising that the spectra of the Hamiltonians-superpartners are (almost) the same. It is important to note, the difference in the spectra of the superpartner Hamiltonians in the ground state signals the exact supersymmetry. Restoring the complete equivalence between two spectra corresponds to the supersymmetry breaking \cite{Witten:1981nf}.

One may wonder, does the presence of Supersymmetry turn out to be so important? We know that experimentally Supersymmetry is still a hidden symmetry of the Nature. However, the fact of absence of superpartners in the realm of particle physics tells us, first of all, about the breaking the Supersymmetry on the LHS action scale. So that, Supersymmetry (SUSY) may be hidden, and this fact can be illustrated by several examples.

\section{Systems with hidden SUSY}

First example of hidden isospectrality is borrowed from gravitational physics. It is well known that gravitational waves are spin-2 fluctuations over a gravitational background. If, for simplicity, the non-trivial background is chosen to be that of a Schwarzschild black hole, it is determined by the so-called red-shift factor $f(r)$. In the linear approximation, dynamics of spin-2 fluctuations $h_s$ over the Schwarzschild gravitational background is determined by a Schr\"odinger-type equation \cite{Regge:1957td,Zerilli:1970se} 
\be
\left[\frac{\partial^2}{\partial r_\ast^2}+\omega^2-V_s(r)\right]h_s=0
\la{RWZeq}
\ee
with the Wheeler coordinate $r_\ast\in(-\infty,+\infty)$; $s=\pm2$. The effective potential of the axial perturbations over the background metric (linearly polarized gravitational waves) comes as follows:
\be
V_{+2}\left(r\right)=-\frac{3f\left(r\right)\partial_rf\left(r\right)}{r}+l(l+1)\frac{f(r)}{r^2} .
\la{V+2}
\ee
For the circularly polarized gravitational waves the effective potential becomes
\be
V_{-2}\left(r\right)=\frac{2f(r)}{r^3}\frac{9M^3+3c^2Mr^2+c^2\left(1+c\right)r^3+9M^2c r}{{(3M+c r)}^2},\quad 
c=\frac{l(l+1)}{2}-1 .
\la{V-2}
\ee
Apparently, in eqs. \rf{V+2}-\rf{V-2} $l$ are integers, starting from $l=2$.

If we insert into eqs. \rf{V+2} and \rf{V-2} the red-shift factor for the Schwarzschild black hole, $f\left(r\right)=1-r_+/r$, we encounter the difference between the effective potentials from \rf{V+2} and \rf{V-2}. However, the effective Hamiltonians are (almost) isospectral, that can be find from the analysis of the effective potential shapes. (See Figure 2.) There are a lot of debates on the origin of such an isospectrality. It is shown for simple backgrounds \cite{Glampedakis:2017rar}; for more exotic configuration of gravitational field the approach fails \cite{Li:2023ulk}.

\begin{figure}[ht]
\begin{center}
\includegraphics[width=3.3in]{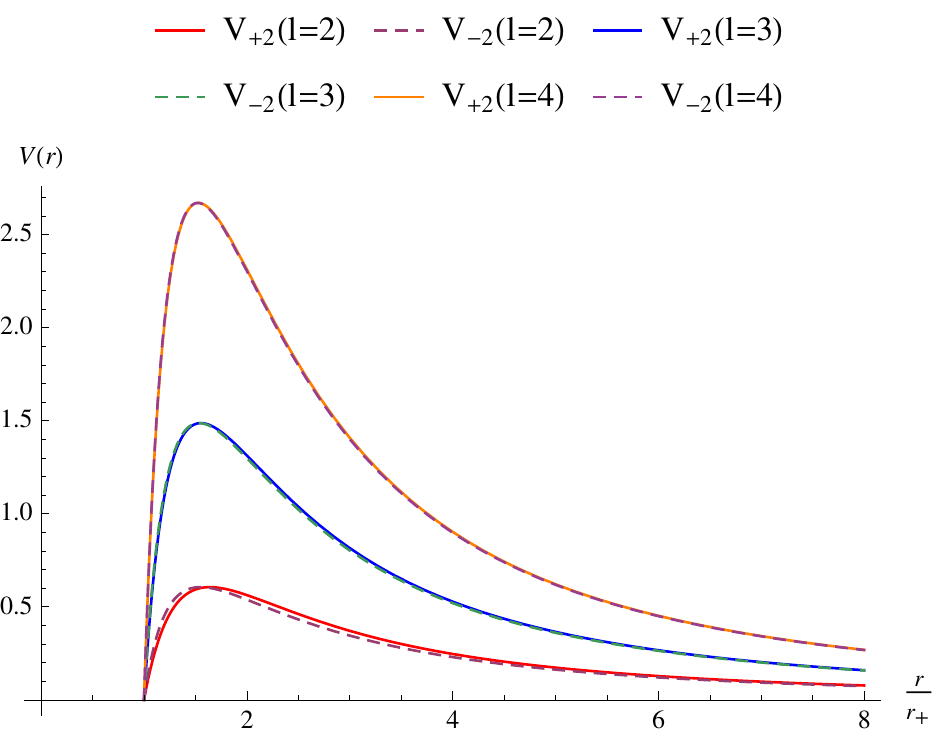}
  \caption{The shapes of effective potentials for axial and polar spin-2 perturbations over the Schwarzschild background. Borrowed from Ref. \cite{Arslanaliev:2021vzm}.}\label{Fig2}
  \end{center}
\end{figure}

Another example of dealing with physical systems with hidden SUSY comes from Quantum Optics. Quantum Optics by itself describes the natural interaction of bosonic (photons) and fermionic (electrons) subsystems of a medium. In the space of parameters of the light-matter interaction it may arise the Bose/Fermi Duality, that transforms, under some specific conditions, into SUSY \cite{Tomka:2015sr}.
Following \cite{Tomka:2015sr}, let’s consider the generalized Rabi model, which is determined by the following Hamiltonian describing a 2-level system interacting with a monochromatic wave:
\be
H=\hbar\omega a^\dag a+\frac{\Delta}{2}\sigma_z+g_1\left(a^\dag\sigma_-+a\sigma_+\right)+g_2\left(a^\dag\sigma_++a\sigma_-\right) .
\la{HRabi}
\ee
In \rf{HRabi}, $\Delta$ is the levels gap; $\omega$ is the boson field frequency; $a^\dag$ and $a$ are the bosonic ladder operators; and, finally, $g_{1,2}$ are arbitrary constants of the light-matter interaction.

In framework of Quantum Optics, the generalized model with Hamiltonian (12) describes the dipole interaction of a monochromatic wave with the bi-level emitter. In the limit of zero-valued constants of interaction, the considered generalized model turns into the Jaynes-Cummings model. When two constants are the same, we get the Rabi model. If one of the interaction constants is small (say, $g_2\rightarrow 0$), the near-resonant consideration (at $\omega\sim \Delta$) corresponds to the rotating-wave approximation (RWA). In the strong interaction constants regime both interaction terms (co- and contra-rotating) should be taken into account.

As it has been proved in \cite{Tomka:2015sr}, SUSY is a symmetry of the generalized Rabi model under the following condition:
\be
g_1^2-g_2^2=\Delta\cdot \omega .
\ee
Then, the supercharge is realized by $4\times 4$ matrix
\be
Q=\left(
\begin{array}{cc}
0&\hat{q}\\0&0\\
\end{array}\right),\qquad 
\hat{q}=\left(
\begin{array}{cc}
g_1/\sqrt\omega&\sqrt\omega\, a\\
\sqrt\omega\, a&g_2/\sqrt\omega\\
\end{array}
\right) . 
\ee
As usual for SUSY (cf. eq. (9)),
\be
\left\{Q,Q^\dag\right\}=H .
\ee
For $g_1=g_2$, SUSY is realized with $\Delta=0$. And Hamiltonian (12) turns into the standard Harmonic Oscillator with a shift.

The next example comes from Condensed Matter Physics. It turns out that SUSY is a hidden symmetry in a topological insulator with Josephson junctions \cite{Galaktionov:2020}. Such a system can be described by the Bogoliubov-de Gennes equations, which, after the appropriate simplifications (see \cite{Galaktionov:2020} for details), can be reduced to the system of differential equations (here and in what follows we use $\partial_x \equiv d/dx$ and $\partial^2_x \equiv d^2/dx^2$)
\be
\left(E+iv\partial_x\right)f(x)+\Delta(x)\phi(x)=0,\qquad
\left(E-iv\partial_x\right)\varphi(x)+\Delta^*(x)f(x)=0 .
\la{BdGeq}
\ee
In \rf{BdGeq}, $E$ is the energy of moving with the velocity $v$ in the positive $x$ direction spinning mode; $\Delta(x)$ is the complex energy gap. It is easy to transform this system of equations in the equation of Witten’s Supersymmetric Quantum Mechanics \cite{Witten:1981nf}
\be
E^2\psi\left(x\right)=\left(-v^2\partial_x^2+{\hat{W}}^2(x)+iv\sigma_z\frac{\partial\hat{W}(x)}{\partial x}\right)\psi(x),\quad 
\sigma_z=\left(
\begin{array}{cc}
1&0\\
0& -1
\end{array}
\right),
\ee
where $\psi(x)$ is the spinning mode state
\be
\psi\left(x\right)=\left(
\begin{array}{c}
f(x)\\
\varphi(x)\\
\end{array}
\right)
\ee
and the {\it superpotential} $\hat{W}(x)$ is given by
\be
\hat{W}(x)=\left(
\begin{array}{cc}
0&\Delta(x)\\
\Delta^*(x) &0
\end{array}
\right) .
\ee
This series of examples could be continued by notable systems with hidden SUSY in Quantum Mechanics \cite{DHoker:1983zea}, nuclear physics \cite{Iachello:1980av} and mesoscopy \cite{Andreev:1987}.

\section{Exactly-solvable models of SQM with multi-well potentials}

In this part of the review, we will focus on the supersymmetric generalization of the Mielnik \cite{Mielnik:1984jmp} construction and its extension to N=4 Supersymmetric Quantum Mechanics \cite{Pashnev:1986zg,Berezovoi:1988tq,Berezovoi:1991xc}.

First of all, let’s describe the technique behind the deformation of the potential shape. Looking at Figure 1, one may notice that the original and the paired Hamiltonians are different in spectra by just one level. And if this new additional level has the energy less than the ground state energy of the original Hamiltonian, it defines the vacuum state of the paired Hamiltonian with new potential (6). To keep the coincident part of the both Hamiltonians spectra, the shape of the new potential shall be changed. The level of deformation depends on the interplay between the unfixed parameters that naturally arise in this scheme upon the definition of new, non-normalized, wave function $\varphi_\epsilon(x)$, corresponding to the new ground state with the factorization energy $\epsilon$. For instance, the Harmonic Oscillator potential describes the system with one potential well;  its deformation may potentially form another well. So that, by controlled adding an additional level, we are able to get the (almost) isospectral quantum mechanical system with two-well potential. However, the clear impact of Quantum Mechanics in this case will consists in possible {\it tunneling effects} between the wells of the paired to the Harmonic Oscillator potential. Further, by adding a new level, lower than the ground level of the Hamiltonian $H_1$, one can form another, paired to $H_1$, Hamiltonian $H_2$ with another potential, the shape of which will be different from the potential $V_1$. Here, the interplay between parameters of the solutions may result as in a two-well as well as in a three-well potential of different (symmetric or asymmetric) shapes. In the latter case, one is able to study more complex tunneling effects that provides a lot of possibilities to control quantum-mechanical processes.

Let us to be more specific and to consider the appearance of the controlling parameters from logic of the construction of supersymmetric Hamiltonians. 

Consider to this end the factorization of a static 1D Hamiltonian $H_+=-\partial^2_x+V_+(x)$ by two operators ({\it supercharges}) $Q^\dag$, $Q$ (analogs of operators $A$ and $A^\dag$ in Section 2):
\be
H_+=Q^\dag Q,\qquad Q^\dag=-\partial_x+\fr12 \partial_x W(x),\quad Q=\partial_x +\fr12 \partial_x W(x) .
\la{H+}
\ee
The {\it superpotential} $W(x)$ \cite{Witten:1981nf} (analog of the function $\beta(x)$ of Section 2) is determined from the Riccati equation
\be
\left(\fr12 \pa_x W(x) \right)^2\mp \fr12 \pa^2_x W(x)=V_\pm (x) ,
\la{RicW}
\ee
where $V_-$ is the potential of the paired to $H_+$ Hamiltonian $H_-=-\pa^2_x+V_-(x)=QQ^\dagger$. The isospectrality of Hamiltonians $H_\pm$ is realized modulo the $H_+$ ground state (the spectrum of the $H_-$ Hamiltonian begins with the first excited level of $H_+$)
\be
\Psi_0(x)=N_0\, e^{-\fr12 W(x)},\qquad E_0=0.
\la{Psi0H+}
\ee
Note, that the zero energy value in the ground state of the initial Hamiltonian is the hallmark of supersymmetry: any of supersymmetric models have a non-negative energy. And if two paired by the supercharges Hamiltonians have the spectrums different just by the ground state, the supersymmetry is exact. 

Now, let's figure out the appearance of a free additional parameter in the approach. Suppose to this end, there exists the factorization of $H_-$ by other supercharges
\be
H_-=\tilde{Q}\tilde{Q}^\dag,\qquad \tilde{Q}=\pa_x+f(x),\quad \tilde{Q}^\dag=-\pa_x+f(x) .
\la{H-}
\ee
From $H_-=-\pa^2_x+V_-(x)$ it comes
\[
f^2(x)+\pa_x f(x)=\left(\fr12 \pa_x W(x) \right)^2+ \fr12 \pa^2_x W(x) .
\]
And if $W(x)$ is known, $f(x)$ is determined by \cite{Mielnik:1984jmp,Kampen:1971}
\be
f(x)=\fr12 \pa_x W(x)-\fr{e^{-W(x)}}{\lambda+\int_x^\infty dz\,e^{-W(z)}}\equiv \fr12 \pa_x W(x)-\f(x) ,
\la{fWdef}
\ee
with $\f(x)$ to be a general solution, with an integration constant $\lambda$, to the equation
\be
\pa_x \f(x)-\f^2(x)+\f(x)\pa_x W(x)=0 .
\la{phieq}
\ee

What will happen when we will construct the Hamiltonian $\tilde{H}_+=\tilde{Q}^\dagger \tilde{Q}$? Explicitly,
\be
\tilde{H}_+ =\tilde{Q}^\dag \tilde{Q}=-\pa^2_x+V_++2\pa_x \f(x),\quad \f(x)=\fr{e^{-W(x)}}{\lambda+\int_x^\infty dz\,e^{-W(z)}},
\la{tilH+}
\ee
and we have constructed the new potential
\be
\tilde{V}_+=V_++2\pa_x \f(x) .
\la{tilV+}
\ee
Apparently, $\tilde{H}_+ \ne H_+$ since the potential $\tilde{V}_+$ in no way equal to the potential $V_+$. However, the spectra of $\tilde{H}_+$ and $H_+$ are the same {\it modulo the ground state}. 

Indeed, by construction, the Hamiltonians $H_+$ and $\tilde{H}_+$ becomes (almost) isospectral, via their parent Hamiltonian $H_-$. Definitely, one can write
\be
\tilde{H}_+\tilde{Q}^\dag=\left(\tilde{Q}^\dag \tilde{Q} \right)\tilde{Q}^\dag=\tilde{Q}^\dag \left(\tilde{Q}\tilde{Q}^\dag \right)=\tilde{Q}^\dag H_- ,
\la{intertwin1}
\ee
and this relation is similar to the one of intertwining relations \rf{intertwin}. According to \rf{intertwin1}, the eigenstates of $\tilde{H}_+$ are determined by $|\tilde{\y}^{(+)}_i\rangle=\tilde{Q}^\dag|\y^{(-)}_i\rangle$, where $|\y^{(-)}_i\rangle$ are the eigenstates of $H_-$. So that, the energies (eigenvalues of $\tilde{H}_+$) will be given by $E_i$, $i=1,\dots,N$ (in the case of finite dimensional and discrete spectrum). However, the ground state $\tilde{Q}|\tilde{\y}^{(+)}_0\rangle=0$ is absent in the spectrum of $\tilde{H}_+$ and has to be added by hands \cite{Mielnik:1984jmp}. In the coordinate representation, the corresponding wave function $\tilde{\Psi}_0$ is
\be
\tilde{\Psi}_0=\tilde{N}_0 \,e^{-\fr12 W(x)} e^{\int_0^x dz\,\f(z)} .
\la{tilPsi0}
\ee
One may straightforwardly check that $\tilde{H}_+ \tilde\Psi_0=0$. So that the spectra (eigenvalues) of $\tilde{H}_+$ and $H_+$ become identical after adding the ground state $ \tilde\Psi_0$.

The appearance of the integration constant $\lambda$ in the potential $\tilde{V}_+$ and in the ground state wave function $\tilde{\Psi}_0$ becomes important for engineering the potential shape. Indeed, by developing the above-mentioned method in the framework of N=4 SQM, we can use both the ground state wave function and the whole spectrum of states in the construction of isospectral Hamiltonians. Moreover, we can use for this purpose even non-normalized wave functions of the starting Hamiltonian, which will correspond to addition of new levels with energies smaller than the ground state energy in the starting Hamiltonian spectrum. (See \cite{Berezovoj:2011ng,Berezovoj:2012ad,Berezovoj:2020uwm} for details.)  This leads to a change of $\phi(x)$, and hence of the integration constant $\lambda$ in \rf{tilH+}. Different values of $\lambda$ define different shapes of the potential: one-well or multi-well, symmetric or asymmetric with respect to the reflection operation. All these properties provide ample opportunities to study quantum mechanical processes and even to control them \cite{Berezovoj:2011ng,Berezovoj:2012ad,Berezovoj:2020uwm}.

As an example of such controlling let us present results of modelling a wave package behavior in a symmetric and a non-symmetric two-well potentials for isospectal Hamiltonians, Figure 3 and Figure 4. Due to the difference in probabilities of the ground and first and second excitation levels in different wells of symmetric and asymmetric potentials, the probability flow of wave packages prepared with the corresponding energies essentially varies. It models the behavior of a {\it quantum diode}. 

\begin{figure}[ht]
\begin{center}
\includegraphics[width=2.3in]{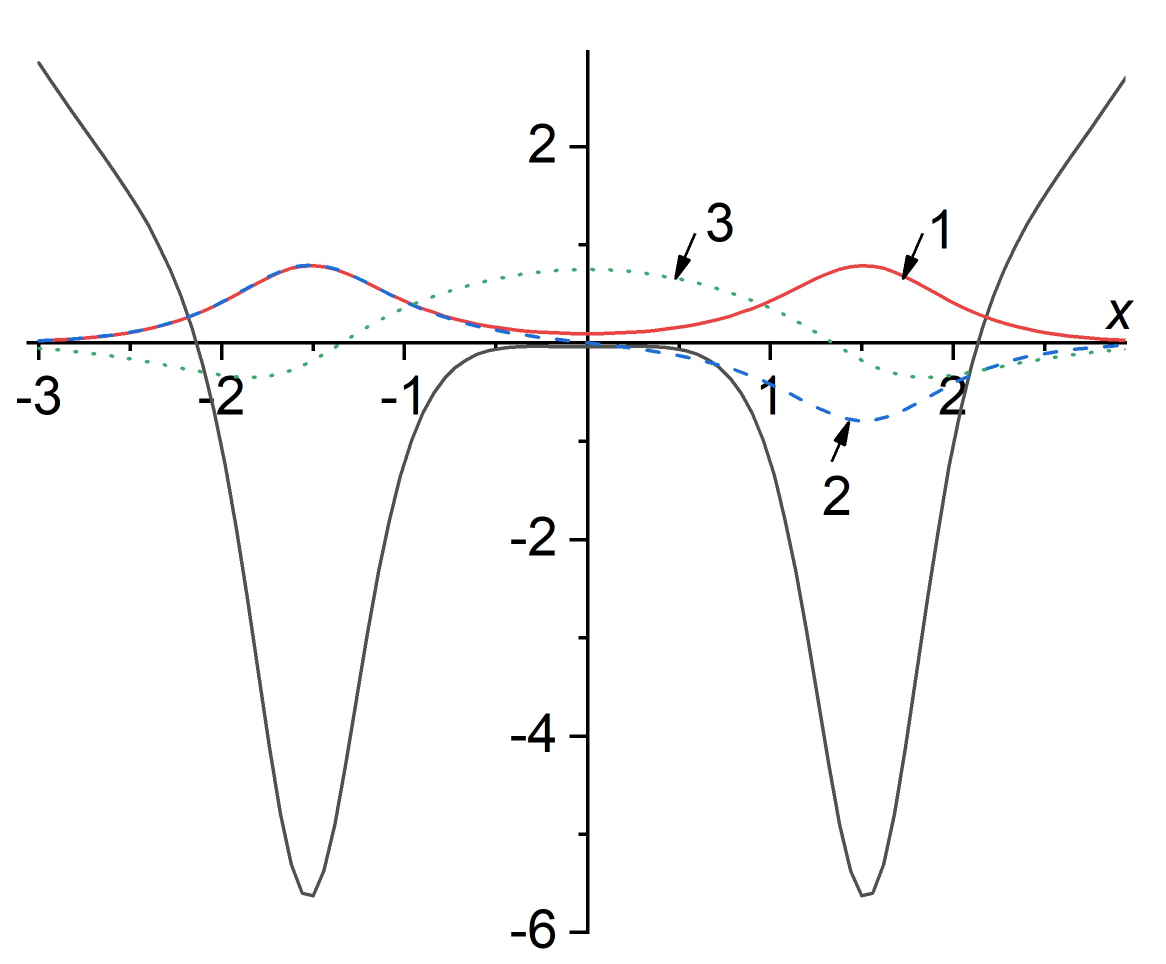}
  \caption{The wave functions of the first three levels in different wells of a symmetric two-well potential; numeration starts with the ground level. See Ref. \cite{Berezovoj:2020uwm} for details.}
\label{Fig3}
  \end{center}
\end{figure}

\begin{figure}[ht]
\begin{center}
\includegraphics[width=2.3in]{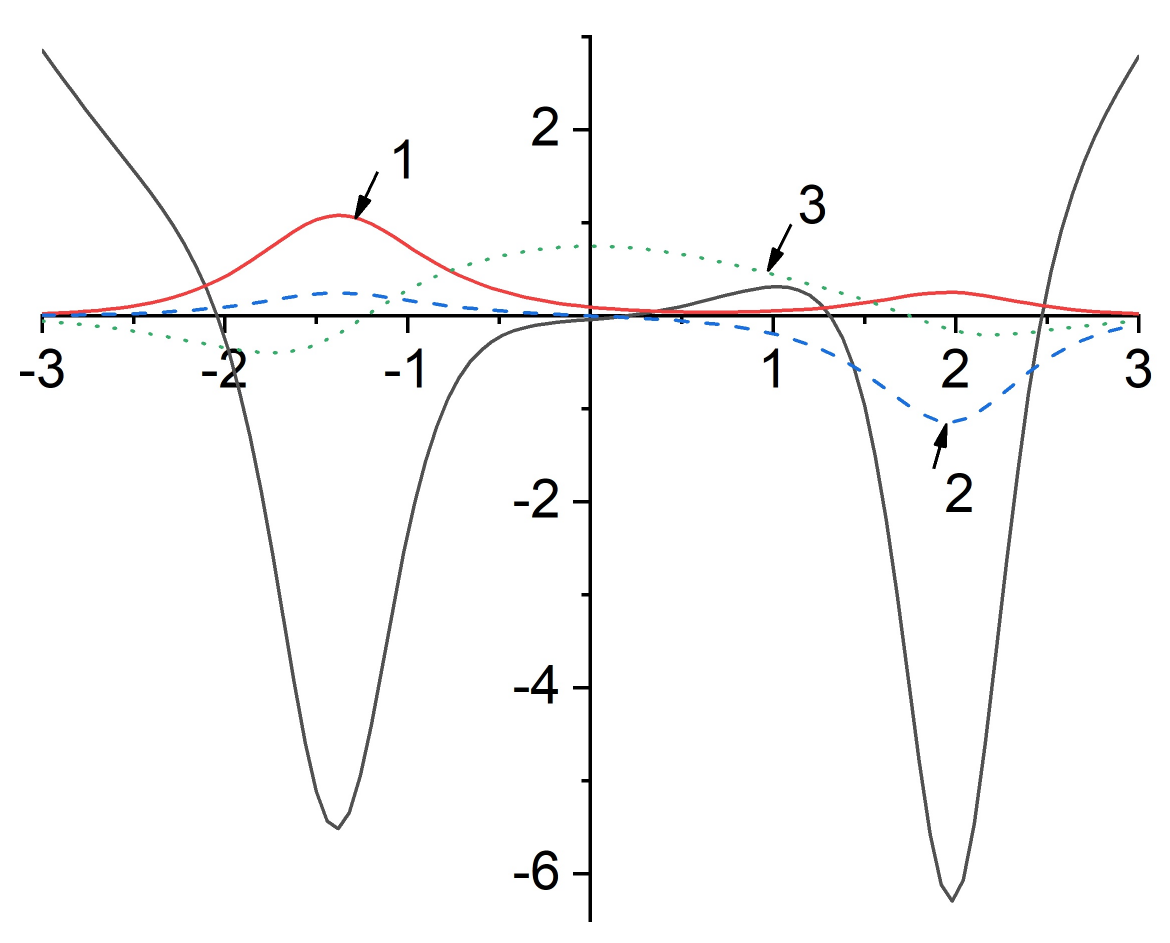}
  \caption{The wave functions of the first three levels in different wells of an asymmetric two-well potential. See Ref. \cite{Berezovoj:2020uwm} for details.}
\label{Fig4}
  \end{center}
\end{figure}

When the number of wells in the isospectral Hamiltonian potential becomes equal to three, the situation becomes more complicated. Here, with tunneling process, one can model properties of a quantum transistor, with different values of ``current'' flow in different wells, see Figure 5.

\begin{figure}[ht]
\begin{center}
\includegraphics[width=2.3in]{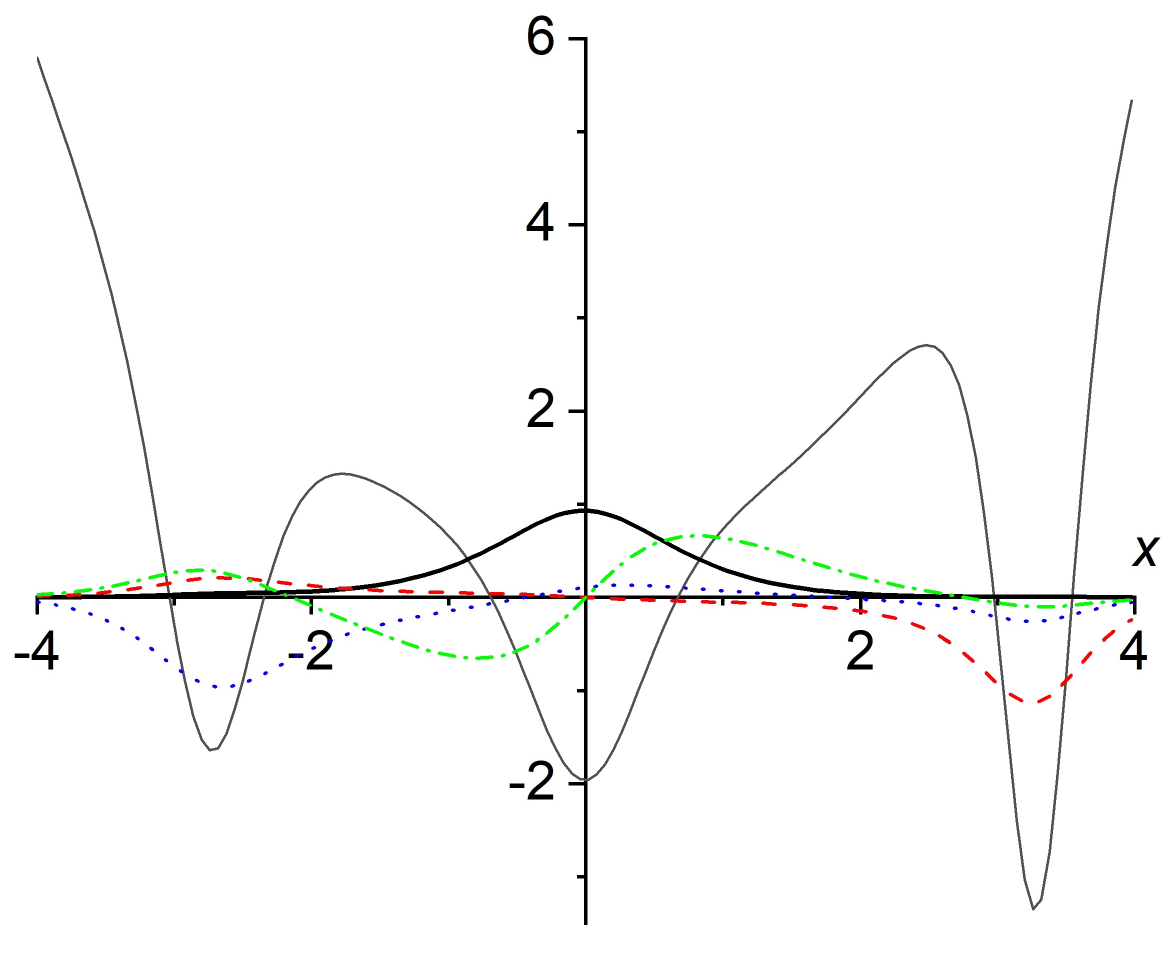}
  \caption{The wave functions of the first three levels in different wells of an asymmetric three-well potential. See Ref. \cite{Berezovoj:2020uwm} for details.}
\label{Fig5}
  \end{center}
\end{figure}

One can generalized the description with adding the temporal dependence. For instance, it could be an external periodic driving force that, in real devices, is reproduced by a laser EM field. The interplay of parameters in the extended by the external field frequency set allows one to reach the phenomenon of the so-called Coherent Tunneling Destruction (CTD) \cite{Grossmann:1991zz}, when without changing the quantum character of the system it becomes possible to localize the initial wave packet in one of the wells. As an illustration of the CTD, in Figure 6 we present the data of numerical simulations borrowed from Ref. \cite{Berezovoj:2020uwm}.

\begin{figure}[ht]
\begin{center}
\includegraphics[width=3.in]{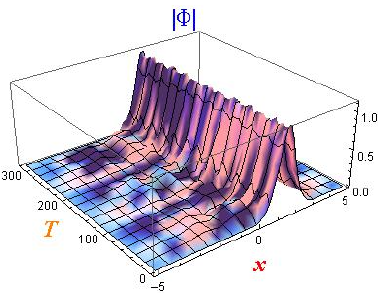}
  \caption{Numerical simulation of evolution of the Gaussian wave packet $\mathrm{\Phi}(x,T)$ in the external periodic driving force. We refer the reader to \cite{Berezovoj:2020uwm} for details.}
\label{Fig6}
  \end{center}
\end{figure}

Hence, the formalism of Supersymmetric Quantum Mechanics turns out to be helpful in modelling different processes such as tunneling, particles flow, diffusion and so on, in the controllable manner. The latter is achieved by the exactly-solvable character of the models in hands, when the Green’s functions are constructed out the explicitly known analytic expressions for the wave functions. Thus, evolution in such systems becomes, if not deterministic, then very predictable.

\section{CPT/PT-invariant Hamiltonians correspondence}

As we have noted in Section 2, the spectra of paired Hamiltonians are connected by the intertwining relations \rf{intertwin}. Their role becomes even more important for non-stationary Supersymmetric Quantum Mechanics, as well as in the case of quantum description of open systems. The following section is devoted to the supersymmetric Lindbladian dynamics. Here we focus on a non-trivial correspondence between CPT and PT invariant models. 

First, let's recall the generalization of the intertwining relations \rf{intertwin1} to the non-stationary SQM: 

\be
\left(i\pa_t -H_- \right)\tilde{Q}=\tilde{Q}\left(i\pa_t-\tilde{H}_+\right).
\la{intertwinNS}
\ee
$H_-$ is the paired to $H_+$ and $\tilde{H}_+$ Hamiltonian (cf. Section 4). For the non-stationary case, $H_+$ becomes the part of the complete Schr\"odinger operator acting on a time-dependent wave function
\be
\left(H_+-i\pa_t \right)\y^{(+)}(x,t)=0,\qquad H_+=-\pa^2_x+V_+(x,t).
\la{SchroNS}
\ee
In the general case, the potential $V_+$ also depends on the time parameter. The way of $H_+$ factorization is implemented as in eq. \rf{H+}, $H_+=Q^\dag Q$, but supercharges $Q$ and $Q^\dag$ now include a time-dependent superpotential $W(x,t)$. The Hamiltonian $H_-$ is realized as $H_-=QQ^\dag$. 

Just as before (cf. equation \rf{H-}), we will require another factorization of $H_-$ in terms of $\tilde{Q}$ and $\tilde{Q}^\dag$ supercharges, which, as now required, must be 
\cite{Bagrov:1990,Bagrov:1991,Bagrov:1995,Bagrov:1996jpa,Samsonov:1996}
\be
\tilde{Q}=l(t)\left(\pa_x+f(x,t) \right),\qquad f(x,t)=\fr12 \pa_x W(x,t)-\f(x,t).
\la{tilQt}
\ee
The functions $l(t)$ and $\f(x,t)$ are generally complex-valued functions, and this fact will be important in what follows. And the use of the intertwining relation \rf{intertwinNS} makes it possible to derive the relation \cite{Bagrov:1990,Bagrov:1991} between the new and the old potentials,
\be
\tilde{V}_+(x,t)=V_+(x,t)+2\pa_x \f(x,t)-i\pa_t \ln l(t).  
\la{VtilCSE}
\ee
The function $\f(x,t)$ is restricted to satisfy 
\be
\pa_x \left[f^2(x,t)-\pa_x f(x,t)-\tilde{V}_+(x,t) \right]-i\pa_t f(x,t)=0,
\la{feqCSE}
\ee
and, if we will require the real-valued spectrum of $\tilde{H}_+$, 
the following constraints should hold:
\be
\pa_t \ln |l(t)|^2=-4 \pa_x \left[ \Im{\mathrm{m}}\, \f(x,t)\right] \qquad \leadsto \qquad \pa^2_x \left[ \Im{\mathrm{m}}\, \f(x,t)\right]=0 .
\la{constraints}
\ee
From the relations \rf{constraints} it becomes clear that for a non-trivial $l(t)$ it is required a complex-valued function $\f(x,t)$.

Now let us associate the stationary Hamiltonian \rf{tilH+} with its time-dependent cousin. Recall, we have an arbitrary integration constant $\lambda$ in the construction of $\tilde{H}_+$ (cf. eqs. \rf{tilH+}). However, {\it changing the $\lambda$ does not affect the spectrum}, affecting the potential and wave functions instead. Therefore, we can instantly change the $\lambda$ in its value, within its range, determined by relation 
\be
\lambda+\int_x^\infty dz \,e^{-W(x)}\ne 0 ,
\la{lambda}
\ee
and still have the steady spectrum. Hence, we can relate any instant $\lambda$ changing to some unique value of the time parameter. Smooth changes will form a function of time $\lambda(t)$ after that \cite{Berezovoj:2024hhg}. 

The so introduced time-dependence of $\lambda$ induces the time-dependence of $\phi(x)$ function of \rf{fWdef} (see \cite{Berezovoj:2024hhg}):
\be
\f(x,t)=\fr{e^{-W(x)}}{\lambda(t)+\int^\infty_x dz\,e^{-W(z)}}.
\la{phitdef}
\ee
Then, the time dependence of the $\tilde{H}_+$ ground state wave function,
\be
\tilde{\Psi}_0(x,t)=\tilde{N}_0 e^{-\fr12 W(x)}e^{\int^x_0 dz\,\f(z,t)},
\la{Psi0tilt}
\ee
can be treated as a time-dependent geometric phase, i.e.,
\be
\tilde{\Psi}_0(x,t)=\tilde{N}_0 e^{-\fr12 W(x)}e^{i\Phi(x,t)},\quad 
i\Phi(x,t)=\int^x_0 dz\,\f(z,t).
\la{Psi0tilt1}
\ee
Wave functions with time-dependent geometric phase naturally appear in quantum systems with time-dependent boundary conditions. (See, e.g., Ref. \cite{Pronin:1991}.) And interesting consequence of introducing time-dependent boundary conditions is introducing the field-compensator in time direction (a vector field with one non-trivial component, to make the formalism close to the standard theory of gauge fields).

To find the Hamiltonian $\tilde{H}_+$ paired to a {\it stationary} Hamiltonian $H_+$,
\be
\left(H_+ -i\pa_t \right) \y(x,t)=0, \qquad H_+=-\pa^2_x+V_+(x),
\la{NSE}
\ee
with the {\it stationary} wave functions $\y(x,t)=e^{-i\ve t} \y_\ve(x)$ and independent on time potential $V_+(x)$ in the situation with time-dependent integration function $\lambda(t)$, one needs to form an analog of the covariant derivative in time direction: $D_t=\pa_t+A_t $. Then, the {\it non-stationary} Hamiltonian $\tilde{H}_+$ obeys the complete Schr\"odinger equation \cite{Berezovoj:2024hhg} 
\be
\left( \tilde{H}_+-iD_t \right) \Psi(x,t)=0,\qquad \tilde{H}_+=-\pa^2_x+\tilde{V}_+(x,t) ,
\la{NSEn}
\ee
and two (almost)isospectral potentials $V_+(x)$ and $\tilde{V}_+(x,t)$ are related to each other as
\be
\tilde{V}_+(x,t)=V_+(x)+2\pa_x \phi(x,t)+\fr{i\pa_t \lambda}{\lambda(t)+\int^\infty_x dz\, |\Psi_0(z)|^2}\,.
\la{tilVS2NS}
\ee
The compensator field is given by
\be
A_t(t)=\fr{\pa_t \lambda(t)}{\lambda(t)+\int^\infty_x dz\, |\Psi_0(z)|^2} ,
\la{At}
\ee
so that the paired non-stationary Hamiltonian $\tilde{H}_+$ becomes complex-valued! In general, the Hermitian condition is not preserved, the energy is not conserved, so that such a Hamiltonian may be used for the description of either an open system or a system with dissipation. (See \cite{Berezovoj:2024hhg} for a brief discussion on possible applications of complex-valued non-stationary Hamiltonians, and Appendix therein.) 

It is important to note that dealing with a real-valued potential to get real-valued eigenvalues of the Hamiltonian is an excess requirement \cite{Baye:1996ytz,Bender:1998ke,Bender:1998gh}. Instead, one may consider complex-valued Hamiltonians with real spectrum, which after \cite{Bender:1998ke,Bender:1998gh} are called as PT-invariant Hamiltonians. The requirement to have a real-valued potential $\tilde{V}_+(x,t)$ leads to the following relation between the function $\phi(x,t)$ and the ``compensator'' field:
\be
\Re{\mathrm{e}} A_t(x,t)=2\pa_x \left(\Im{\mathrm{m}}\, \phi(x,t) \right).
\la{AKrel}
\ee
Hence, in the case of PT-invariant Hamiltonians, $\phi(x,t)$ should be complex-valued to supply the non-triviality of the compensator $A_t(x,t)$. This requirement essentially narrows the class of physical systems to describe. More on the application of complex-valued potentials with complex-valued spectra can be found in Ref. \cite{Moiseyev:Book}.

\section{Supersymmetric Quantum Mechanics for open systems}

As we have noted in previous sections, the spectra of paired Hamiltonians and their potentials are connected by the intertwining relations \rf{intertwin}, \rf{intertwin1}, \rf{intertwinNS}. They become main relations for the developed in  \cite{Bagrov:1990,Bagrov:1991,Bagrov:1995,Bagrov:1996jpa,Samsonov:1996} non-stationary SQM,  as well as in the polynomial (nonlinear, higher derivative) formulation of Supersymmetric Quantum Mechanics \cite{Andrianov:1993md,Andrianov:1994aj,Andrianov:1994mk,Andrianov:2012vk}. The latter approach has been used in extending the SQM to the description of open systems with exchanging the energy with the environment. We have noted that complex-valued Hamiltonians can be potentially used to this end. The gold standard in the description of open quantum systems is the application of the Franke-Gorini-Kossakowski-Lindblad-Sudarshan \cite{Franke:1976tx,Gorini:1975nb,Lindblad:1975ef,Lindblad:1976wv} (FGKLS) equation. The supersymmetrization of the FGKLS equation was considered in Ref. \cite{Andrianov:2019wso}, and here we closely follow this approach.   

The quantum states of an open system are mixed. Thus, one needs to use the density matrix formalism, in which the density matrix operator of a sub-system evolves according to
\be
\fr{\pa \rho}{\pa t}=-i\left(H_{\mathrm{eff}} \,\rho-\rho\, H^\dag_{\mathrm{eff}}\right)+L[\rho] .
\la{Leq}
\ee  
Here $H_{\mathrm{eff}}$ is the effective Hamiltonian (which is not Hermitian in the case), and $L[\rho]$ is the so-called Lindbladian
\be
L[\rho]=\sum_j \left(A_j \rho A^\dag_j-\fr12 \left[\rho, A^\dag_j A_j \right] \right). 
\la{Lop}
\ee
The $A_j$ operators are unspecified, and they may be used to model the influence of the measurement devices on a quantum system. The Lindbladian operator is introduced to evolve pure states into mixed states. So that $A_j$ operators are naturally restricted to provide increasing the entropy 
\be
S={\mathrm {Tr}} \left(-\rho \log \rho \right)
\la{Sdef}
\ee
upon the evolution of the system. 

To construct the supersymmetric version of the FGKLS equation \rf{Leq}, let's introduce the supermultiplet of two paired Hamiltonians
\be
\cal{H}=\left(
\begin{array}{cc}
H_+ & 0\\
0 & H_-
\end{array}
\right)=\left(
\begin{array}{cc}
-\pa^2_x+V_+(x) &0\\
0 & -\pa^2_x+V_-(x)
\end{array}
\right).
\la{SH}
\ee
Following \cite{Andrianov:2019wso} we suppose the super-Hamiltonian to be independent on time.

Isospectrality of $H_+$ and $H_-$ follows from the intertwining relations \rf{intertwin} (replacing $(H_0, H_1)$ with $(H_+,H_-)$, respectively) 
\be
A H_+=H_- A,\qquad H_+ A^\dag=A^\dag H_- .
\la{intertwinL}
\ee
The supercharges $Q$ and $Q^\dag$ are constructed out the $A$, $A^\dag$ operators:
\be
Q=\left(
\begin{array}{cc}
0 & A^\dag \\
0 & 0
\end{array}
\right),\qquad
Q^\dag=\left(
\begin{array}{cc}
0 & 0 \\
A & 0
\end{array}
\right) .
\la{QQdagdef}
\ee
Then, the intertwining relations \rf{intertwinL} reflect the conservation of supercharges during the evolution of the system:
\be
[{\cal{H}},Q]=[{\cal{H}},Q^\dag]=0.
\la{HQrel}
\ee
The full superalgebra is closed by adding the following anti-commutator of supercharges,
\be
\{Q,Q^\dag\}={\cal P}({\cal H}),
\la{QQH}
\ee 
where, in dependence on the realization of $A$ and $A^\dag$ by first or higher order derivative operators, ${\cal P}({\cal H})$ is a linear or polynomial function of the super-Hamiltonian, respectively. (See \cite{Andrianov:2019wso} and Refs. therein for details.)

Similarly to the super-Hamiltonian $\cal{H}$ we can introduce the super-density matrix
\be
\mathfrak{P}=\left(
\begin{array}{cc}
\rho_+ &0\\
0 & \rho_-
\end{array}
\right).
\la{rho}
\ee 
Then, the Schr\"odinger-Liouville equation (for $L[\mathfrak{P}]=0$) is
\be
\fr{\pa}{\pa t}\,\mathfrak{P}=-i[{\cal H},\mathfrak{P}],
\la{rhoeq}
\ee
and we have to find the analog of intertwining relations \rf{intertwinL} for super-partners $\rho_+$ and $\rho_-$. We can use the standard definition of the mean value to this end. For the super-Hamiltonian it becomes
\be
\langle {\cal P}({\cal H}) \rangle={\mathrm{Tr}} \left({\cal P}({\cal H}) \,\mathfrak{P} \right)
={\mathrm{Tr}} \left(\sqrt{{\cal P}({\cal H})}\, \mathfrak{P} \,\sqrt{{\cal P}({\cal H})} \right)
={\mathrm{Tr}} (Q\, \mathfrak{P} \,Q^\dag+Q^\dag\, \mathfrak{P} \, Q ),
\la{Hmean}
\ee
and the proposed by the authors of Ref. \cite{Andrianov:2019wso} intertwining relations are
\be
Q^\dag\,\mathfrak{P}\, \sqrt{{\cal P}({\cal H})}=\sqrt{{\cal P}({\cal H})} \,\mathfrak{P}\,Q^\dag,\qquad Q\,\mathfrak{P}^\dag\,\sqrt{{\cal P}({\cal H})}=\sqrt{{\cal P}({\cal H})}\,\mathfrak{P}^\dag\,Q .
\la{intertwinrho}
\ee
Analogous relations can be used for the Lindbladian operator with generally non-Hermitian $A_j$. For the super-density matrix the Hermiticity condition still holds: $\mathfrak{P}=\mathfrak{P}^\dag$.

The generalization of \rf{Lop} to the SQM is apparent: we have to introduce
\be
\mathfrak{A}_j=\left(
\begin{array}{cc}
A_{j+} &0\\
0 & A_{j-}
\end{array}
\right)
\la{Asup}
\ee
and to form, by use of \rf{rho} and \rf{Asup}, the super-Lindbladian operator. The intertwining relations for linear and quadratic in $\mathfrak{A}_{j}$ combinations (cf. \rf{intertwinrho}) are:
\be
Q^\dag\, \mathfrak{A}_j \,\sqrt{{\cal P}({\cal H})}=\sqrt{{\cal P}({\cal H})}\, \mathfrak{A}_j \,Q^\dag,\qquad Q \,\mathfrak{A}^\dag \,\sqrt{{\cal P}({\cal H})}=\sqrt{{\cal P}({\cal H})}\, \mathfrak{A}^\dag\, Q,
\la{intertwinL1}
\ee
\be
Q^\dag\, \sum_j \mathfrak{A}_j^\dag \mathfrak{A}_j \,\sqrt{{\cal P}({\cal H})}=\sqrt{{\cal P}({\cal H})}\, \sum_j \mathfrak{A}_j^\dag \mathfrak{A}_j \,Q^\dag,\qquad Q \,\sum_j \mathfrak{A}_j^\dag \mathfrak{A} \,\sqrt{{\cal P}({\cal H})}=\sqrt{{\cal P}({\cal H})}\, \sum_j \mathfrak{A}_j^\dag \mathfrak{A}\, Q .
\la{intertwinL2}
\ee
Examples of the application of the formalism can be found in Ref. \cite{Andrianov:2019wso}.

\section{Conclusions}

In conclusion, let us recall the main advantages of Supersymmetry that make this approach preferable to others. 

First, as it has been pointed out in one of the underlying work on Supersymmetry \cite{Deser:1976eh}, supersymmetric models are super-renormalizable ones: the UV divergency problem, actual for standard field theories including the Standard Model, is completely absent for them. Unfortunately, our Universe is not supersymmetric; but this fact is about the energy scale on which SUSY is broken. 

Second, application of Supersymmetry as a tool to investigate various physical models in different regimes and on different energy scales has shown its self-consistency and efficiency. We can just cite a few quotes from the modern literature in favor of this claim. For instance, the authors of Ref. \cite{Tomka:2015sr} write: `` … optical and condensed matter systems at the SUSY points can be used for quantum information technology and can open an avenue for quantum simulation of the SUSY field theories.'' In Ref. \cite{Jolie:2019ynl} one can read off: `` … the atomic nucleus $^{195}$Pt represents an excellent example of the dynamical U(6/12) supersymmetry. … certainly the best documented example of the manifestation of dynamical supersymmetry in atomic nuclei.'' As it comes from reading Ref. \cite{Crichigno:2011}, the potential of supersymmetry has not yet been definitively revealed, since the author ``… brings the attention to the role of supersymmetry in quantum computation and quantum information more broadly, a subject much underexplored.'' Furthermore, in our studies \cite{Berezovoj:2011ng,Berezovoj:2012ad,Berezovoj:2020uwm} we have noticed the advantages of using the SQM approach, such as full knowledge of the spectrum and explicit analytical expressions for the wave functions, which makes studies of the evolution of quantum systems simple and straightforward, via the exact Green's function. The Coherent Tunneling Destruction and other tunneling effects in multi-well potentials require within the SQM approach a finite number of states (of the order of ten), while the same numerical simulation for potentials constructed by stitching a harmonic oscillator potential requires consideration of hundreds of states.   

And finally, here we have shown the perspectives of applying the SQM formalism in studying quantum sytems. Especially in cases closer to the real world \cite{Andrianov:2019wso}. Recently, we have extended the approach to include the temporal dependence into the game \cite{Berezovoj:2024hhg}, that made it possible to relate CPT-invariant stationary Hamiltonians to their PT-invariant non-stationary partners. It opens new avenues in investigations of quantum-mechanical models with complex-valued potentials, having more reach structure of physical phenomena, and being applicable to an essentially wide class of physical systems, including open systems. We expect new non-trivial results along the way and will strive to contribute to exciting intertwining Supersymmetry with quantum realm.


\subsection*{Acknowledgments} We express our sincere gratitude to Dmitry Vasilievich Volkov, who introduced us to the magical world of supersymmetry. 
We are also grateful to M.I. Konchatnij for the pleasant collaboration, which resulted in many important new results.  
The work of A.J.N. is partly carried out as a part of the COST initiative CA22113 -- Fundamental challenges in theoretical physics (THEORY-CHALLENGES).

\subsection*{Conflicts of interest} The authors declare no conflict of interest.




\end{document}